\documentclass[12pt]{article}
\usepackage{epsf,amsfonts,hyperref}
\renewcommand{\appendix}[1]{
    \addtocounter{section}{1}
    \setcounter{equation}{0}
    \renewcommand{\thesection}{\Alph{section}}
    \section*{Appendix \thesection\protect\indent #1}
    \addcontentsline{toc}{section}{Appendix \thesection\ \ \ #1}
}
\newcommand\encadremath[1]{\vbox{\hrule\hbox{\vrule\kern8pt
\vbox{\kern8pt \hbox{$\displaystyle #1$}\kern8pt}
\kern8pt\vrule}\hrule}}
\def\enca#1{\vbox{\hrule\hbox{
\vrule\kern8pt\vbox{\kern8pt \hbox{$\displaystyle #1$}
\kern8pt} \kern8pt\vrule}\hrule}}

\newcommand\figureframex[3]{
\begin{figure}[bth]
\hrule\hbox{\vrule\kern8pt
\vbox{\kern8pt \vbox{
\begin{center}
{\mbox{\epsfxsize=#1.truecm\epsfbox{#2}}}
\end{center}
\caption{#3}
}\kern8pt}
\kern8pt\vrule}\hrule
\end{figure}
}
\newcommand\figureframey[3]{
\begin{figure}[bth]
\hrule\hbox{\vrule\kern8pt
\vbox{\kern8pt \vbox{
\begin{center}
{\mbox{\epsfysize=#1.truecm\epsfbox{#2}}}
\end{center}
\caption{#3}
}\kern8pt}
\kern8pt\vrule}\hrule
\end{figure}
}

\renewcommand{\thesection}{\arabic{section}}

\makeatletter
\@addtoreset{equation}{section}
\makeatother
\newtheorem{theorem}{Theorem}[section]
\newtheorem{conjecture}{Conjecture}[section]
\newtheorem{remark}{Remark}[section]
\newtheorem{proposition}{Proposition}[section]
\newtheorem{lemma}{Lemma}[section]
\newtheorem{corollary}{Corollary}[section]
\newtheorem{definition}{Definition}[section]
\def\br{\begin{remark}\rm\small}
\def\er{\end{remark}}
\def\bt{\begin{theorem}}
\def\et{\end{theorem}}
\def\bd{\begin{definition}}
\def\ed{\end{definition}}
\def\bp{\begin{proposition}}
\def\ep{\end{proposition}}
\def\bl{\begin{lemma}}
\def\el{\end{lemma}}
\def\bc{\begin{corollary}}
\def\ec{\end{corollary}}
\def\beaq{\begin{eqnarray}}
\def\eeaq{\end{eqnarray}}
\newcommand{\proof}[1]{{\noindent \bf proof:}\par
{#1} $\square$}

\newcommand{\beq}{\begin{equation}}
\newcommand{\eeq}{\end{equation}}
\newcommand{\bea}{\begin{eqnarray}}
\newcommand{\eea}{\end{eqnarray}}

\renewcommand{\and}{{\qquad {\rm and} \qquad}}

\newcommand{\virg}{{\qquad , \qquad}}

 \newcommand{\Tr}{{\,\rm Tr}\:}

\newcommand{\Res}{\mathop{\,\rm Res\,}}

\newcommand{\td}[1]{{\tilde{#1}}}

\newcommand{\om}{\omega}

\newcommand{\ee}[1]{{{\rm e}^{#1}}}

\newcommand{\C}{{\mathbf C}}
\newcommand{\VV}{{\cal V}}

\newcommand{\Pint}{{\int\kern -1.em -\kern-.25em}}

\renewcommand{\Re}{{\mathrm{Re}}}
\renewcommand{\Im}{{\mathrm{Im}}}

\newcommand{\sn}{{\rm sn}}
\newcommand{\cn}{{\rm cn}}

\newcommand{\ovl}{\overline}

\begin{document}
%=============================Page de titre==============%\date{??}
%\author{Eynard}
%\title{Correlation functions for hermitian random matrices}
%\topmargin .5cm \textheight 21.5cm \textwidth 15.8cm
%\oddsidemargin 0.54cm
%\evensidemargin 0.54cm
\sloppy

%\maketitle

\pagestyle{empty}
\hfill SPhT-T05/241
\addtolength{\baselineskip}{0.20\baselineskip}
\begin{center}
\vspace{26pt}
{\large \bf {Formal matrix integrals and combinatorics of maps}}
\newline
\vspace{26pt}

{\sl B.\ Eynard}\hspace*{0.05cm}\footnote{ E-mail: eynard@spht.saclay.cea.fr }
\vspace{6pt}
Service de Physique Th\'{e}orique de Saclay,\\
F-91191 Gif-sur-Yvette Cedex, France.\\
\end{center}

\vspace{20pt}
\begin{center}
{\bf Abstract}:
\end{center}

This article is a short review on the relationship between convergent matrix integrals, formal matrix integrals, and combinatorics of maps.
%All the results presented here have been developed over the last 30 years, and are not my own results, but a summarised rewriting of physicist's works in a more mathematical presentation.

%-----------------------------ABSTRACT--------------------------------------
%
%Abstract

%\begin{center}

%\end{center}

%\newpage
%\pagestyle{empty}

%\section*{}

%\newpage
\vspace{26pt}
\pagestyle{plain}
\setcounter{page}{1}

%*********************************************************************
%==================== ARTICLE =======================================%*********************************************************************

\section{Introduction}

This article is a short review on the relationship between {\bf convergent matrix integrals}, {\bf formal matrix integrals}, and {\bf combinatorics of maps}.
We briefly summarize results developed over the last 30 years, as well as more recent discoveries.

We recall that formal matrix integrals are {\bf identical} to combinatorial generating functions for maps, and that formal matrix integrals are in general very {\bf different} from convergent matrix integrals.
Both may coincide perturbatively (i.e. up to terms smaller than any negative power of $N$), only for some potentials which correspond to negative weights for the maps, and therefore not very interesting from the combinatorics point of view.

We also recall that both convergent and formal matrix integrals are solutions of the same set of loop equations, and that loop equations do not have a unique solution in general.

Finally, we give a list of the classical matrix  models which have  played an important role in physics in the past decades.
Some of them are now well understood, some are still difficult challenges.

\medskip

Matrix integrals were first introduced by physicists \cite{Wigner}, mostly in two ways:\\
- in nuclear physics, solid state physics, quantum chaos, {\em convergent matrix integrals} are studied for the eigenvalues statistical properties \cite{Mehta, Guhr, BI, Moerbeke:2000}.
Statistical properties of the spectrum of large random matrices show some amazing universal behaviours, and it is believed that they correspond to some kind of ``central limit theorem'' for non independent random variables. This domain is very active and rich, and many important recent progresses have been achieved by the mathematicians community. Universality was proved in many cases, in particular using the Riemann-Hilbert approach of Bleher-Its \cite{BlIt} and Deift Venakides Zhou Mac Laughlin \cite{dkmvz}, and also by large deviation methods \cite{guionet, guionnet2}.\\
- in Quantum Chromodynamics, quantum gravity, string theory, conformal field theory, {\em formal matrix integrals} are studied for their combinatorial property of being generating functions of maps \cite{DGZ}.
This fact was first discovered by t'Hooft in 1974 \cite{thooft:1974}, then further developed mostly by BIPZ \cite{BIPZ} as well as Ambjorn, David, Kazakov \cite{DGZ, Matrixsurf, gross:1991, KazakovGQ, KazakovGQbis}.
For a long time, physicist's papers have been ambiguous about the relationship between formal and convergent matrix integrals, and many people have been confused by those ill-defined matrix integrals.
However, if one uses the word ``formal matrix integral'', many physicist's results of the 80's till now are perfectly rigorous, especially those using loop equations.
Only results regarding convergency properties were non rigorous, but as far as combinatorics is concerned, {\bf convergency is not an issue}.

The ambiguity in physicist's ill-defined matrix integrals started to become obvious when E. Kanzieper and V. Freilikher \cite{Kanz}, and later Brezin and Deo in 1998 \cite{BrezinDeo} tried to compare the topological expansion of a formal matrix integral derived from loop equations, and the asymptotics of the convergent integral found with the help of orthogonal polynomials. The two results did not match. The orthogonal polynomial's method showed clearly that the convergent matrix integrals had no large N power series expansion (it contained some $(-1)^N$). The origin of this puzzle has now been understood \cite{BDE}, and it comes from the fact that formal matrix integrals and convergent matrix integrals are different objects in general.

%\smallskip

%So, I decided to make this short review to define unambiguously ``what physicists are talking about'', i.e. the notion of formal matrix integrals.

\bigskip

This short review is only about combinatoric properties of formal matrix integrals. Matrix models is a very vast topic, and many important applications, methods and points of view are not discussed here.
In particular, critical limits (which include asymptotics of combinatoric properties of maps), the link with integrable systems, with conformal field theory, with algebraic geometry, with orthogonal polynomials, group theory, number theory, probabilities and many other aspects, are far beyond the scope of such a short review.

\section{Formal matrix integrals}

In this section we introduce  the notion of formal matrix integrals of the form:
\beq
Z_{\rm formal}(t) = \int_{\rm formal} dM_1\, dM_2\, \dots \, dM_p\,\,\,  \ee{-{N\over t} \left( \sum_{i,j=1}^p {C_{ij}\over 2}\Tr{ M_i M_j}  \,\, - N V(M_1,\dots,M_p)\right) }
\eeq
The idea is to formally expand the exponential $\ee{{N^2\over t}V}$ in powers of $t^{-1}$, and compute the Gaussian integral for each term.
The result is a formal series in powers of $t$.
So, let us define it precisely.

\medskip

\bd
$Q$ is an invariant non-commutative monomial of $M_1,\dots,M_p$, if $Q=1$ or if $Q$ is of the form:
\beq
\ovl{Q} = \prod_{r=1}^R {1\over N}\Tr( W_r )
\eeq
where each $W_r$ is an arbitrary word written with the alphabet $M_1,\dots,M_p$.
$Q$ is the equivalence class of $\ovl{Q}$ under permutations of the $W_r$'s, and cyclic permutations of the letters of each $W_r$.

The degree of $Q$ is the sum of lengths of all $W_r$'s.

Invariant non-commutative polynomials of $M_1,\dots,M_p$ are complex finite linear combinations of monomials:
\beq
V = \sum_{Q} t_{Q}\, Q \virg t_Q\in \C
\eeq
The degree of a polynomial is the maximum degree of its monomials.

They are called invariant, because they are left unchanged if one conjugates all matrices $M_i\to U M_i U^{-1}$ with the same invertible matrix $U$.
\ed
Invariant polynomials form an algebra over $\C$.

Let $V(M_1,\dots,M_p)$ be an arbitrary invariant polynomial of degree $d$ in $M_1,\dots,M_p$, which contains only monomials of degree at least $3$.

\bp
Let $C$ be a $p\times p$ symmetric positive definite matrix, then the following Gaussian integral
\beq
A_k(t) = {
\int_{H_N\times\dots\times H_N} dM_1\, dM_2\, \dots \, dM_p\,\,\, {N^{2k}\,t^{-k}\over k!} V^k\,\, \ee{-{N\over 2t}\Tr \sum_{i,j=1}^p C_{ij} M_i M_j }
\over
\int_{H_N\times\dots\times H_N} dM_1\, dM_2\, \dots \, dM_p\,\,\,  \ee{-{N\over 2t}\Tr \sum_{i,j=1}^p C_{ij} M_i M_j }
}
\eeq
where $dM_i$ is the usual Lebesgue ($U(N)$ invariant) measure on the space of hermitian matrices $H_N$, is absolutely convergent and has the following properties:

$A_k(t)$ is a polynomial in $t$, of the form:
\beq\label{Aktpol}
A_k(t) = \sum_{k/2 \leq j \leq kd/2-k} A_{k,j}\, t^j
\eeq

$A_k(t)$ is a Laurent polynomial in $N$.

$A_0(t)=1$.

\ep

\proof{$A_0=1$ is trivial. Let $d=\deg V$. Since $V$ is made of monomials of degree at least $3$ and at most $d$, then $V^k$ is a sum of invariant monomials whose degree $l$ is between $3k$ and $dk$.
According to Wick's theorem, the Gaussian integral of a monomial of degree $l$ is zero if $l$ is odd, and it is proportional to $t^{l/2}$ if $l$ is even.
Since $3k\leq l\leq dk$ we have:
\beq
0\leq k/2 \leq l/2-k\leq dk/2 -k
\eeq
Thus $A_k(t)$ is a finite linear combination of positive integer powers of $t$, i.e. it is a polynomial in $t$, of the form of eq.\ref{Aktpol}.

The matrix size $N$'s dependence comes in several ways. First there is the factor $N^{2k}$.
The matrix size also appears in the matrix products (each matrix product is a sum over an index which runs from $1$ to $N$), in the traces (it means the first and last index of a matrix product have to be identified, thus there is a Kroenecker's $\delta_{ij}$ of two indices).
And after Gaussian integration over all matrix elements, the Wick's theorem pairings result in $N^{-l/2}$ times some product of Kroenecker's $\delta$ of pairs of indices (times some elements of the matrix $C^{-1}$ which are independent of $N$).
The matrix indices thus appear only in sums and $\delta's$, and the result of the sum over indices is an integer power of $N$.
Thus, each $A_k(t)$ is a finite sum (sum for each monomial of $V^k$, and the Gaussian integral of each monomial is a finite sum of Wick's pairings) of positive or negative powers of $N$, i.e. a Laurent polynomial in $N$.
}

\bd\label{defZj}
The formal matrix integral $Z_{\rm formal}(t)$
 is defined as the formal power series:
\beq
Z_{\rm formal}(t) = \sum_j Z_j\,t^j
\virg
Z_j = \sum_{k=0}^{2j} A_{k,j}
\eeq
and each $Z_i$ is a Laurent polynomial in $N$.
Notice that $Z_0=1$.

\ed

By abuse of notation, $Z_{\rm formal}(t)$ is often written:
\beq
Z_{\rm formal}(t) =
{\int_{\rm formal} dM_1\, dM_2\, \dots \, dM_p\,\,\,  \ee{-{N\over t} \left( \sum_{i,j=1}^p {C_{ij}\over 2}\Tr{ M_i M_j}  \,\, - N V(M_1,\dots,M_p)\right) }
\over
\int_{H_N\times\dots\times H_N} dM_1\, dM_2\, \dots \, dM_p\,\,\,  \ee{-{N\over 2t}\Tr \sum_{i,j=1}^p C_{ij} M_i M_j }
}
\eeq
but it does not mean that it has anything to do with the corresponding convergent (if it converges) integral.
In fact, the integral can be absolutely convergent only if $\deg(V)$ is even and if the $t_Q$ corresponding to the highest degree terms of $V$ have a negative real part.
But as we shall see below, the relevant case for combinatorics, corresponds to all $t_Q$'s positive, and in that case, the formal integral is NEVER a convergent one.

\bd\label{defFj}
The formal free energy $F_{\rm formal}(t)$ is defined as the formal log of $Z_{\rm formal}$.
\beq
F_{\rm formal}(t) = \ln{(Z_{\rm formal}(t))} = \sum_{j} F_j\, t^j
\eeq
We have $F_0=0$.
Each $F_j$ is a Laurent polynomial in $N$.
\ed

\subsection{Combinatorics of maps}

Recall that an invariant monomial is a product of terms, each term being the trace of a word in an alphabet of $p$ letters.
Thus, an invariant monomial is given by:\\
$\bullet$ the number $R$ of traces, ($R-1$ is called the crossing number of $Q$),\\
$\bullet$ $R$ words written in an alphabet of $p$ letters.

The $R$ words can be permuted together, and in each word the letters can be cyclically permuted.
We label the invariant monomials by the equivalence classes of those permutations.

Another graphical way of coding invariant monomials is the following:

\bd
To each invariant monomial $Q$ we associate biunivoquely a $Q-gon$ (generalized polygon) as follows: \\
$\bullet$ to each word we associate an oriented polygon (in the usual sense), with as many edges as the length of the word, and whose edges carry a ``color'' between $1$ and $p$, given by the corresponding letter in the word.\\
$\bullet$ the $R$ words are glued together by their centers on their upper face (in accordance with the orientation), so that they make a surface with $R-1$ crossings.\\
$\bullet$ $R-1$ which is the number of traces minus one (i.e. one trace corresponds to a crossing number zero), is called the crossing number of the $Q$-gon.\\
$\bullet$ The degree $\deg(Q)$ of the $Q$-gon is the total number of edges (sum of lengths of all words).\\
$\bullet$ to $Q$-gon we associate a symmetry factor $s_Q={\rm \#Aut}(Q)$ which is the number of symmetries which leave $Q$ unchanged.

\begin{figure}[bth]
\hrule\hbox{\vrule\kern8pt
\vbox{\kern8pt \vbox{
\begin{center}
{\mbox{\epsfxsize=5truecm\epsfbox{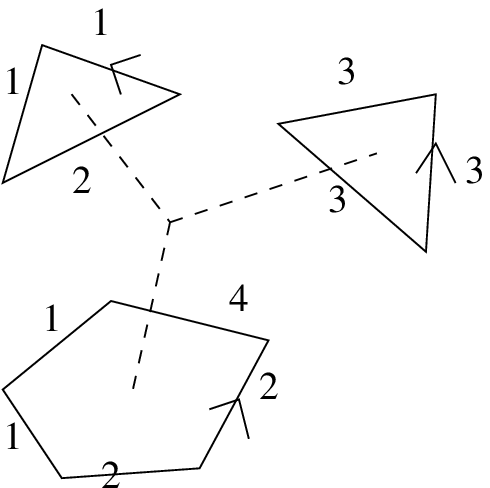}}}
\end{center}
\caption{The invariant monomial $Q=N^{-3}\Tr(M_1^2 M_2) \Tr(M_3^3) \Tr(M_2^2 M_4 M_1^2)$ of degree $11$, is represented by 2 triangles and one pentagon glued together by their center. The dashed lines mean that the 3 centers should actually be seen as only one point. Its symmetry factor is $s_Q=3$ because we can perform 3 rotations on the triangle of color (3,3,3).\label{figpolygons}}
}\kern8pt}
\kern8pt\vrule}\hrule
\end{figure}

\ed
An example is given in fig.\ref{figpolygons}.
Notice that we allow a $Q$-gon to be made of polygons with possibly one or two sides.
We will most often call the $Q$-gons polygons. The usual polygons are $Q$-gons with no crossing i.e. $R=1$.

\bd
Let $p\geq 1$ and $d\geq 3$ be given integers.
Let ${{\cal Q}}_{d,p}$ be the set of all invariant monomials $Q$ (or polygons) of degree $3\leq \deg(Q)\leq d$.\\
${{\cal Q}}_{d,p}$ is clearly a finite set.
\ed

\bd
Let ${\cal S}_{l,d,p}$ be the set of oriented discrete surfaces such that \#edges-\#Qgons+\#crossings = $l$, and obtained by gluing together polygons (belonging to ${{\cal Q}}_{d,p}$) by their sides (following the orientation).
The edges of the polygons carry colors among $p$ possible colors (thus each edge of the surface, is at the border of two polygons, and has two colors, one on each side).

Let $\overline{{\cal S}}_{l,d,p}$ be the subset of ${\cal S}_{l,d,p}$ which contains only connected surfaces.

\smallskip
Such surfaces are also called ``maps''.
\ed

\bp
${\cal S}_{l,d,p}$ is a finite set.
\ep

\proof{Indeed, since all polygons of ${{\cal Q}}_{d,p}$ have at least 3 sides, we have \#edges $\geq {3\over 2}$\#polygons,
and thus \#edges $\leq 2l$ and \#polygons $\leq 4l /3 $, and thus the number of discrete surfaces in ${\cal S}_l$, is finite.
We can give a very large bound on \#${\cal S}_{l,d,p}$.
We have:
\beq
\#{\cal S}_{l,d,p} \leq (4dl/ 3)^p \, (2dl/3)! \,(4dl/3)! \leq (4dl/ 3)^p \, (2dl)!
\eeq

}

\bt (t'Hooft 1974 and followers)\label{ththooft}:
If the potential $V$ is an invariant polynomial given by
\beq
V = \sum_{Q\in {{\cal Q}}_{d,p}}\, {t_Q\over s_Q} \, Q
\eeq
then:
\beq
Z_l  = \sum_{S\in {\cal S}_{l,d,p}} {1\over \# {\rm Aut}(S)}\,N^{\chi(S)}\,\,\prod_Q t_Q^{n_Q(S)}\, \prod_{i,j} \left((C^{-1})_{i,j}\right)^{E_{i,j}(S)/2}
\eeq
\beq
F_l  = \sum_{S\in \overline{{\cal S}}_{l,d,p}} {1\over \# {\rm Aut}(S)}\,N^{\chi(S)}\,\,\prod_Q t_Q^{n_Q(S)}\, \prod_{i,j} \left((C^{-1})_{i,j}\right)^{E_{i,j}(S)/2}
\eeq
where $Z_l$ and $F_l$ were defined in def.\ref{defZj}, def.\ref{defFj}, and where:\\
$\bullet$ $n_Q(S)$ is the number of $Q$-gons in $S$, \\
$\bullet$ $E_{ij}(S)$ is the number of edges of $S$ which glue two polygons whose sides have colors $i$ and $j$,\\
$\bullet$ $\chi(S)$ is the Euler characteristic of $S$.\\
$\bullet$ ${\rm Aut}(S)$ is the group of automorphisms of $S$, i.e. the group of symmetries which leave $S$ unchanged. $\#{\rm Aut}(S)$ is called the symmetry factor.\\

In other words,
$Z_l$ is the formal generating function which enumerates discrete surfaces of ${\cal S}_{l,d,p}$, according to their Euler characteristics, their number of polygons, number of edges according to their color, number of crossings...
$F_l$ is the formal generating function  for the same surfaces with the additional condition that they are connected.

\begin{figure}[bth]
\hrule\hbox{\vrule\kern8pt
\vbox{\kern8pt \vbox{
\begin{center}
{\mbox{\epsfxsize=8truecm\epsfbox{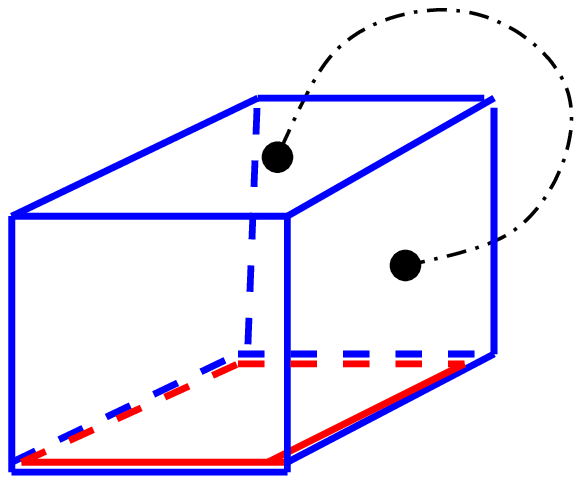}}}
\end{center}
\caption{If $V={t_4\over 4} {1\over N}\Tr M_1^4 +{\td{t}_4\over 4} {1\over N}\Tr M_2^4+ {t_{4,4}\over 32 N^2} (\Tr M_1^4)^2$, the above surface contributes to the term $N^0\,\,t_4^3\, \td{t}_4\, t_{4,4}\, (C^{-1})_{1,1}^4\,(C^{-1})_{1,2}^4\,$.
Indeed $M_1$ is represented in blue, $M_2$ in red, so that $\Tr M_1^4$ corresponds to blue squares, $\Tr M_2^4$ corresponds to red squares, and $(\Tr M_1^4)^2$ corresponds to pairs of blue squares glued at their center. Its Euler characteristic is $\chi=0$ (it is a torus), and this surface has no automorphism (other than identity), i.e. \#Aut=1. It corresponds to $l=7$.\label{figsurface}}
}\kern8pt}
\kern8pt\vrule}\hrule
\end{figure}

\et
An example is given in fig. \ref{figsurface}.

\proof{ It is a mere application of Feynman's graphical representation of Wick's theorem\footnote{Although Feynman's graphs are sometimes regarded as non-rigorous, let us emphasize that it is only when Feynman's graphs and Wick's theorem are used for functional integrals that they are non-rigorous. Here we have finite dimensional Gaussian integrals, and Wick's theorem and Feynman's graphs are perfectly rigorous.}.

The only non-trivial part is the power of $N$, because $N$ enters not only in the weights but also in the variables of integrations.
It was the great discovery of t'Hooft to recognize that the power of $N$ is precisely the Euler characteristics.
Let us explain it again.

First, $V^k$ is decomposed into a combination of monomials:
\beq
V=\sum_Q {t_Q\over s_Q}\, Q
\virg
V^k = \sum_G T_{k,G} \, G
\eeq
where $G$ is a product of $Q$'s, and $T_{k,G}=\sum_{Q_1\cup Q_2\cup\dots\cup Q_k=G} t_{Q_1} t_{Q_2}\dots t_{Q_k}\,{1\over \prod_i s_{Q_i}}$.
$G$ is thus a collection of polygons, some of them glued by their centers. So far the polygons of $G$ are not yet glued by their sides to form a surface.
Remember that each $Q$ carries a factor $N^{-R}$ if $Q$ is made of $R$ traces.

Then, for each $G$, the Gaussian integral is computed by Wick's theorem and gives a sum of Wick's pairings, i.e. it is  the sum over all possible ways of pairing two $M$'s, i.e. it is the sum over all possible ways of gluing together all polygons by their sides, i.e. corresponds to the sum over all surfaces $S$.
The pairing $<M_{i\, ab} M_{j\, cd}>$ of the $(a,b)$ matrix element of $M_i$ and the $(c,d)$ matrix element of $M_j$ gives a factor:
\beq
<M_{i\, ab} M_{j\, cd}>= {t\over N}\, (C^{-1})_{ij}\,\,\,\delta_{ad}\delta_{bc}
\eeq

The double line notation for pairings (see \cite{DGZ})
allows to see clearly that the sum over all indices is $N^{\hbox{\#vertices}(S)}$.
The total power of $N$ is thus:
\beq
2k-\sum_Q R_Q  +\hbox{\#vertices} - \hbox{\#edges} - \hbox{\#polygons}
\eeq
Now, notice that
\beq
\hbox{\#polygons} = \sum_Q R_Q = k +  \sum_Q (R_Q-1) = k + \hbox{\#crossings}
\eeq
Thus the total power of $N$ is:
\beq
\hbox{\#vertices} - \hbox{\#edges} + \hbox{\#polygons} - 2 \hbox{\#crossings} = \chi
\eeq
which is the Euler characteristic of $S$.

\medskip

We leave to the reader to see in the literature how to find the symmetry factor.
}

\bc
$N^{-2} F_l$ is a polynomial in $N^{-2}$:
\beq
F_l = \sum_{g=0}^{l+1} \, N^{2-2g} \,F_{l,g}
\eeq
\ec
\proof{Indeed the Euler characteristic of connected graphs is of the form $\chi=2-2g$ where $g$ is a positive integer and is the genus (ànumber of handles). The fact that it is a polynomial comes from the fact that $F_l$ is a finite sum.

It is easy to find some bound on $g$.
We have:
\bea
2g-2
&=& -\hbox{\#vertices} + \hbox{\#edges} - \hbox{\#polygons} + 2 \hbox{\#crossings} \cr
&=& -\hbox{\#vertices} + 2l  - \hbox{\#edges} + \hbox{\#polygons}  \cr
\eea
and using $\hbox{\#edges} \geq 3/2 \hbox{\#polygons}$, we have:
\beq
2g-2 \leq  -\hbox{\#vertices} + 2l - {1\over 2} \hbox{\#polygons}  \leq  2l
\eeq

}

\subsection{Topological expansion}

\bd
We define the genus $g$ free energy as the formal series in $t$:
\beq
F^{(g)}(t) = - \sum_{l=0}^{\infty} \, t^{l+2-2g} \,F_{l,g}
\eeq
$F^{(g)}$ is the generating function for connected oriented discrete surfaces of genus $g$.
\ed

Remark: the minus sign in front of $F^{(g)}$ is there for historical reasons, because in thermodynamics the free energy is $-\ln{Z}$.

\medskip

There is no reason a priori to believe that $F^{(g)}(t)$ might have a finite radius of convergence in $t$.
However, for many classical matrix models, it is proved that $\forall g$, $F^{(g)}$ has a finite radius of convergence because it can be expressed in terms of algebraic functions.

There is also no reason a priori to believe  that the $F^{(g)}$'s should be the limits of convergent matrix integrals.
There are many works which prove that the number of terms in $F^{(g)}$ grows with $g$ like $(\beta g)!$ for some $\beta$.
If $t$ and all $t_Q$'s are positive (this is the interesting case for combinatorics and statistical physics because we want all surfaces to be counted with a positive weight), then $F^{(g)}$ is  positive and grows like $(\beta g)!$, and therefore the sum of the topological series CANNOT be convergent (even under Borel resummation).
For negative $t$, it is only an asymptotic series, and at least in some cases, it can be made convergent using Borel transforms.

\section{Loop equations}

Loop equations is the name given to Schwinger-Dyson equations in the context of random matrices \cite{Wadia}. The reason for that name is explained below.
Let us recall that Schwinger-Dyson equations  are in fact nothing but integration by parts, or the fact that the integral of a total derivative vanishes.

In particular, in the context of convergent matrix integrals we have:
\bea\label{loopeqSDderivtotal}
0&=&
\sum_{i<j}\, \int_{H_N^p} dM_1\, dM_2\, \dots \, dM_p\,\,\, \left({\partial \over \partial \Re M_{k\, i,j}}-i{\partial \over \partial \Im M_{k\, i,j}}\right) \cr
&& \left( (G(M_1,\dots,M_p)+G(M_1,\dots,M_p)^\dagger)_{ij} \,   \ee{-{N\over t} \left( \sum_{i,j=1}^p {C_{ij}\over 2}\Tr{ M_i M_j}  \,\, - N V(M_1,\dots,M_p)\right) }\right) \cr
&& +\sum_{i}\, \int_{H_N^p} dM_1\, dM_2\, \dots \, dM_p\,\,\, {\partial \over \partial M_{k\, i,i}} \cr
&& \left( (G(M_1,\dots,M_p)+G(M_1,\dots,M_p)^\dagger)_{ii} \,   \ee{-{N\over t} \left( \sum_{i,j=1}^p {C_{ij}\over 2}\Tr{ M_i M_j}  \,\, - N V(M_1,\dots,M_p)\right) }\right) \cr
\eea
where $G$ is any non commutative polynomial, and $k$ is an integer between $1$ and $p$.

Therefore, Schwinger--Dyson equations for matrix integrals give relationships between expectation values of invariant polynomials.
Namely:
\beq\label{loopeqgeneral}
{N\over t} \left<\Tr\left(\left(\sum_j C_{kj} M_j - N D_k(V)\right) G\right)\right>
= \left<K_k(G)\right>
\eeq
where $D_k$ is the non commutative derivative with respect to $M_k$, and $K_k(G)$ is some invariant polynomial which can be computed by the following rules:

$\bullet$ Leibnitz rule
\bea
&& K_k( A(M_1,\dots,M_k,\dots,M_p) B(M_1,\dots,M_k,\dots,M_p)) \cr
&=& \left( K_k(A(M_1,\dots,M_k,\dots,M_p) B(M_1,\dots,m_k,\dots,M_p))\right)_{m_k\to M_k} \cr
&& + \left( K_k(A(M_1,\dots,m_k,\dots,M_p) B(M_1,\dots,M_k,\dots,M_p))\right)_{m_k\to M_k}
\eea

$\bullet$ split rule
\bea
&& K_k( A(M_1,\dots,m_k,\dots,M_p) M_k^l B(M_1,\dots,m_k,\dots,M_p)) \cr
&=& \sum_{j=0}^{l-1} \Tr( A(M_1,\dots,m_k,\dots,M_p) M_k^j)\Tr(M_k^{l-j-1}B(M_1,\dots,m_k,\dots,M_p))
\eea

$\bullet$ merge rule
\bea
&& K_k( A(M_1,\dots,m_k,\dots,M_p) \Tr(M_k^l B(M_1,\dots,m_k,\dots,M_p))) \cr
&=& \sum_{j=0}^{l-1} \Tr( A(M_1,\dots,m_k,\dots,M_p) M_k^j B(M_1,\dots,m_k,\dots,M_p) M_k^{l-j-1})
\eea

$\bullet$ no $M_k$ rule
\beq
K_k( A(M_1,\dots,m_k,\dots,M_p))=0
\eeq

Since each rule allows to strictly decrease the partial degree in $M_k$, this set of rules allows to compute $K_k(G)$ for any $G$.

For any $G$ and any $k$ we get one loop equation of the form eq.\ref{loopeqgeneral}.

\bd
The formal expectation value of some invariant polynomial $P(M_1,\dots,M_k)$ is the formal power series in $t$ defined by:
\beq
A_{k,P}(t) = {
\int_{H_N\times\dots\times H_N} dM_1\, dM_2\, \dots \, dM_p\,\,\, {N^{2k}\,t^{-k}\over k!} P\,V^k\,\, \ee{-{N\over 2t}\Tr \sum_{i,j=1}^p C_{ij} M_i M_j }
\over
\int_{H_N\times\dots\times H_N} dM_1\, dM_2\, \dots \, dM_p\,\,\,  \ee{-{N\over 2t}\Tr \sum_{i,j=1}^p C_{ij} M_i M_j }
}
\eeq

$A_{k,P}(t)$ is a polynomial in $t$, of the form:
\beq\label{AkPtpol}
A_{k,P}(t) = \sum_{\deg(P)/2+k/2 \leq j \leq \deg(P)/2+kd/2-k} A_{k,P,j}\, t^j
\eeq
and we define the following quantity
\beq
A_{P,j}(t) = \sum_{k=0}^{2j-\deg P} A_{k,P,j}
\eeq
and the formal series
\beq
A_P(t) = \sum_{j=\deg P/2}^\infty A_{P,j}\, t^j
\eeq
Again, each $A_{k,P}$, $A_{k,P,j}$, $A_{P,j}$ is a Laurent polynomial in $N$.

The formal expectation value of $P$ is defined as:
\beq
\left< P(M_1,\dots,M_k) \right>_{\rm formal} = {A_P(t)\over Z_{\rm formal}(t)}
\eeq
where the division by $Z_{\rm formal}$ is to be taken in the sense of formal series, and it can be performed since $Z_{\rm formal}(t)=\sum_{j=0}^\infty Z_j t^j$ with $Z_0=1$.

\ed

The formal expectation value is often written by abuse of notations
\beq
<P>_{\rm formal} =
{\int_{\rm formal} dM_1\, dM_2\, \dots \, dM_p\,\,\, P\,\, \ee{-{N\over t} \left( \sum_{i,j=1}^p {C_{ij}\over 2}\Tr{ M_i M_j}  \,\, - N V(M_1,\dots,M_p)\right) }
\over
\int_{\rm formal} dM_1\, dM_2\, \dots \, dM_p\,\,\, \ee{-{N\over t} \left( \sum_{i,j=1}^p {C_{ij}\over 2}\Tr{ M_i M_j}  \,\, - N V(M_1,\dots,M_p)\right) }
}
\eeq

\bt
The formal expectation values of the formal matrix integral satisfy the same loop equations as the convergent matrix integral ones, i.e. they satisfy eq.\ref{loopeqgeneral} for any $k$ and $G$.
\et

\proof{
It is clear that Gaussian integrals, and thus formal integrals satisfy eqs.\ref{loopeqSDderivtotal}.
The remainder of the derivation of loop equations for convergent integrals is purely algebraic, and thus it holds for both convergent and formal integrals.}

\smallskip

On a combinatoric point of view, loop equations are the generalisation of Tutte's equations for counting planar maps \cite{Tutte1, Tutte2}.
This is  where the name ``loop equations'' comes from: indeed, similarly to theorem.\ref{ththooft}, formal expectation values of traces are generating functions for open discrete surfaces with as many boundaries as the number of traces (and again the power of $N$ is the Euler characteristic of the surface). The boundaries are ``loops'', and the combinatorical interpretation of Schwinger-Dyson equations is a recursion on the size of the boundaries, i.e. how to build discrete surfaces by gluing loops \`a la Tutte \cite{Tutte1, Tutte2}.

\medskip

Notice that in general, the loop equations {\bf don't have a unique solution}.
One can find a unique solution only with additional constraints not contained in the loop equations themselves.
Thus, the fact that both convergent and formal matrix models obey the same loop equations does not mean they have to coincide.
Many explicit examples where both do not coincide have been found in the literature.
It is easy to see on a very simple example that Schwinger--Dyson equations can't have a unique solution: consider the Airy function
$\int_\gamma \ee{{1\over 3}t^3-tx}\,dt$ where $\gamma$ is any path in the complex plane, going from $\infty$ to $\infty$ such that the integral is convergent. There are only two homologicaly independent choices of $\gamma$ (one going from $+\infty$ to $\ee{2i\pi/3}\infty$ and one from $+\infty$ to $\ee{-2i\pi/3}\infty$). Schwinger-Dyson equations are: $<n t^{n-1} + t^{n+2} - x t^n>=0$ for all $n$. It is clear that loop equations are independent of the path $\gamma$, while their solution clearly depends on $\gamma$.

\bt
The formal matrix integral is the solution of loop equations with the following additional requirements:

- the expectation value of every monomial invariant is a formal power series in $N^{-2}$.

- the $t\to 0$ limit of the expectation value of any monomial invariant of degree $\geq 1$ vanishes:
\beq
\mathop{{\rm lim}}_{t\to 0} <Q>=0 \qquad {\rm if}\,\, Q\neq 1
\eeq

\et
\proof{The fact that all expectation values have a formal $N^{-2}$ expansion follows from the construction.
The fact that $\mathop{{\rm lim}}_{t\to 0} <Q>=0 \qquad {\rm if}\,\, Q\neq 1$, follows from the fact that we are making a formal Taylor expansion at small $t$, near the minimum of the quadratic part, i.e. near $M_i=0$, $i=1,\dots,p$.}

However, even those requirements don't necessarily provide a unique solution to loop equations.
Notice that there exist formal solutions of loop equations which satisfy the first point (there is a formal $N^{-2}$ expansion), but not the second point ($\mathop{{\rm lim}}_{t\to 0} <Q>=0$).
Those solutions are related to so-called ``multicut'' solutions, they also have a known combinatoric interpretation, but we don't consider them in this short review (see for instance \cite{eynhabilit, BDE} for examples).

\bigskip

There is a conjecture about the relationship between different solutions  of loop equations:
\begin{conjecture}
The convergent matrix integral (we assume $V$ to be such that the integral is convergent, i.e. the highest $t_Q$'s have negative real part)
\beq
Z_{\rm conv} = \int_{H_N^p} dM_1\, dM_2\, \dots \, dM_p\,\,\,  \ee{-{N\over t} \left( \sum_{i,j=1}^p {C_{ij}\over 2}\Tr{ M_i M_j}  \,\, - N V(M_1,\dots,M_p)\right) }
\eeq
is a finite linear combination of convergent formal solutions of loop equations (i.e. a formal solution of loop equations $\ln{Z} = - \sum_{g=G}^\infty N^{2-2g} F^{(g)}$, such that the $N^{-2}$ series is convergent.), i.e.
\beq\label{conjecture}
Z_{\rm conv} = \sum_i c_i Z_i \virg \ln{Z_i} = -\sum_{g=0}^\infty N^{2-2g}\,F^{(g)}_i
\eeq

\end{conjecture}

{\bf Hint:}
It amounts to exchanging the large $N$ and small $t$ limits.
First, notice that convergent matrix integrals are usualy defined on $H_N^p$, but can be defined on any "integration path" in the complexified of $H_N^p$, which is $M_N(\C)^p$, as long as the integral is convergent.
The homology space of such contours is of finite dimension (because there are a finite number of variables $p\times N^2$, and a finite number of possible sectors at $\infty$ because the integration measure is the exponential of a polynomial).
Thus, the set of "generalized convergent matrix integrals" defined on arbitrary paths, is a finite dimensional vector space which we note: ${\rm Gen}$. The hermitian matrix integral defined on $H_N^p$ is only one point in that vector space.

Second, notice that every such generalized convergent matrix integral satisfies the same set of loop equations, and that loop equations of type eq.\ref{loopeqSDderivtotal} are clearly linear in $Z$.
Thus, the set of solutions of loop equations is a vector space which contains the vector space of generalized convergent matrix integrals.

Third, notice that formal integrals are solutions of loop equations, and therefore, formal integrals which are also convergent, belong to the vector space of generalized convergent matrix integrals.

Fourth, one can try to compute any generalized convergent matrix integral by small $t$ saddle point approximation (at finite $N$).
In that purpose, we need to identify the small $t$ saddle points, i.e. a set of matrices $(\underline{M}_1,\dots, \underline{M}_p) \in M_N(\C)^p$ such that $\forall i,j,k$ one has:
\beq
{\partial\over \partial M_{k_{i,j}}}\,\left(\sum_{l,m} \Tr ({1\over 2}C_{l,m} M_l M_m) - V(M_1,\dots,M_p)\right)_{M_n=\underline{M}_n} = 0
\eeq
and such that
\beq
\Im \left(\sum_{l,m} \Tr ({1\over 2}C_{l,m} M_l M_m) - V(M_1,\dots,M_p)\right) =\Im \left(\sum_{l,m} \Tr ({1\over 2}C_{l,m} \underline{M}_l \underline{M}_m) - V(\underline{M}_1,\dots,\underline{M}_p)\right)
\eeq
and
\beq
\Re \left(\sum_{l,m} \Tr ({1\over 2}C_{l,m} M_l M_m) - V(M_1,\dots,M_p)\right) \geq \Re \left(\sum_{l,m} \Tr ({1\over 2}C_{l,m} \underline{M}_l \underline{M}_m) - V(\underline{M}_1,\dots,\underline{M}_p)\right)
\eeq
If such a  saddle point exist, then it is possible to replace $\exp{{N^2\over t}V}$ by its Taylor series expansion in the integral and exchange the summation and the integration (because both the series and the integral are absolutely convergent, this is nothing but WKB expansion). This proves that saddle points are formal integrals and at the same time they are generalized convergent integrals, thus they are formal convergent integrals.

The conjecture thus  amounts to claim that saddle points exist, and that there exist as many saddle points as the dimension of ${\rm Gen}$, and that they are independent, so that the saddle points form a basis of $\rm Gen$.

\bigskip

Notice that a linear combination of convergent formal solutions has no $N^{-2}$ expansion in general, and thus the set of convergent formal integrals is not a vector space.

This conjecture is proved for the 1-matrix model with real potential \cite{dkmvz, EML}, and for complex potentials it can be derived from Bertola-Man Yue \cite{marcomo} (indeed, the asymptotics of the partition function follow from those of the orthogonal polynomials).
It is the physical intuition that it should hold for more general cases.
It somehow corresponds to the small $t$ saddle point method. Each saddle point has a WKB expansion, i.e. is a convergent formal solution, and the whole function is a sum over all saddle points.
The coefficients of the linear combination reflect the homology class of the path (here $H_N^p$) on which $Z$ is defined. This path has to be decomposed as a linear combination of steepest descent paths.
The coefficients of that linear combination make the $c_i$'s of eq.\ref{conjecture} (this is the generalisation of the idea introduced in \cite{BDE}).

\medskip

\section{Examples}

\subsection{1-matrix model}

\beq
Z_{\rm formal} = {\int dM \ee{-{N\over t}\Tr(\VV(M))} \over \int dM \ee{-{N\over t}\Tr({C\over 2}M^2)}}
\eeq
where $\VV(M)={C\over 2}M^2 -V(M)$ is a polynomial in $M$, such that $V(M)$ contains only terms with degree $\geq 3$.

For any $\alpha$ and $\gamma$, let us parametrise the potential $\VV$ as:
\beq
\left\{
\begin{array}{l}
x(z)=\alpha+\gamma(z+1/z) \cr
\VV'(x(z)) = \sum_{j=0}^d v_j (z^j+z^{-j})
\end{array}\right.
\eeq
We determine $\alpha$ and $\gamma$ by the following conditions:
\beq
v_0 = 0 \virg
v_1 = {t\over \gamma}
\eeq
i.e. $\alpha$ and $\beta$ are solutions of an algebraic equation and exist for almost any $t$, and they are algebraic functions of $t$.
In general, they have a finite radius of convergence in $t$.
We chose the branches of $\alpha(t)$ and $\gamma(t)$ which vanish at $t=0$:
\beq
\alpha(t=0)=0 \virg \gamma(t=0)=0
\eeq

Then we define:
\beq
y(z) = {1\over 2}\sum_{j=1}^d v_j (z^j-z^{-j})
\eeq
The curve $(x(z),y(z))\, z\in\C$ is called the spectral curve. It is a genus zero hyperelliptical curve $y^2 = {\rm Polynomial}(x)$.
It has only two branch points solutions of $x'(z)=0$ which correspond to $z=\pm 1$, i.e. $x=\alpha \pm 2\gamma$.
$y$ as a function of $x$ has a cut along the segment $[\alpha-2\gamma,\alpha+2\gamma]$.
Notice that we have:
\beq
\Res_{z\to\infty} y dx  = t = - \Res_{z\to 0} y dx
\eeq
\beq
\Res_{z\to\infty} \VV'(x) y dx  = 0 \virg  \Res_{z\to \infty} x \VV'(x) y dx = t^2
\eeq

Then one has \cite{marcoF}:
\bea
F^{(0)}&=&{1\over 2}\left( \Res_{z\to\infty} \VV(x) y dx - t \Res_{z\to\infty} \VV(x) {dz\over z}  -{3\over 2}t^2 -  t^2 \ln{\left(\gamma^2 C\over t\right)}  \right) \cr
&=& {1\over 2}\left( -\sum_{j\geq 1} {\gamma^2\over j} (v_{j+1}-v_{j-1})^2 -{2\gamma t\over j}(-1)^j (v_{2j-1}-v_{2j+1}) -{3\over 2}t^2 - t^2 \ln{(C \gamma^2/t)}   \right) \cr
\eea
and: \cite{chekhovF1,  EKK}
\beq
F^{(1)}= -{1\over 24}\ln{\left({\gamma^2 y'(1)y'(-1)\over t^2} \right)}
\eeq

Expressions are known for all $F^{(g)}$'s and we refer the reader to \cite{eynloop1mat, eynchekhov}.
Those expressions are detailed in the next section about the 2-matrix model, and  one has to keep in mind that the 1-matrix model is a special case of the 2-matrix model with $\VV_1(x) = \VV(x)+{x^2\over 2}$ and $\VV_2(y)={y^2\over 2}$.

As a  fore-taste, let us just show an expression for $F^{(2)}$:
\bea
-2 F^{(2)}
&=& \Res_{z_1\to \pm 1}\Res_{z_2\to \pm 1}\Res_{z_3\to \pm 1} \Phi(z_1) E(z_1,z_2) E(1/z_1,z_3) {1\over (z_2-{1\over z_2})^2} {1\over (z_3-{1\over z_3})^2} \cr
&& + 2\Res_{z_1\to \pm 1}\Res_{z_2\to \pm 1}\Res_{z_3\to \pm 1} \Phi(z_1) E(z_1,z_2) E(z_2,z_3) {1\over ({1\over z_1}-{1\over z_2})^2} {1\over (z_3-{1\over z_3})^2} \cr
&& + 2\Res_{z_1\to \pm 1}\Res_{z_2\to \pm 1}\Res_{z_3\to \pm 1} \Phi(z_1) E(z_1,z_2) E(z_2,z_3) {1\over ({1\over z_1}-{1\over z_3})^2} {1\over ({1\over z_2}-z_3)^2} \cr
\eea
where the residue is first evaluated for $z_3$ then $z_2$ then $z_1$, and
where:
\beq
E(z,z') = {1\over 4\gamma}\,{1\over z' (z-z')(z-{1\over z'}) y(z')}
\eeq
\beq
\Phi(z) = -{1\over 4\gamma}\,{1\over z y(z) (z-{1\over z})}\,\int_{1/z}^z y dx
\eeq

\subsubsection{Example triangulated maps}

Consider the particular case $V(x)={t_3\over 3}x^3$.

Let $T={t t_3^2\over C^3}$, and $a$ be a solution of:
\beq
a-a^3=4T
\eeq
and consider the branch of $a(T)$ which goes to $1$ when $T=0$.
If we parametarize:
\beq
T={t t_3^2\over C^3} = {\sin{3\phi}\over 6\sqrt{3}}
\eeq
we have:
\beq
a = {\cos{({\pi\over 6}-\phi)}\over \cos{({\pi\over 6})}}
\eeq

We have:
\beq
\alpha ={C\over 2t_3}\,(1-a)
\virg
\gamma^2 = {t\over a C}
\eeq
\beq
v_0=0
\virg
v_1= {t\over \gamma}
\virg
v_2=-t_3 \gamma^2
\eeq
which gives:
\beq
F^{(0)} = {5\over 12} t^2 - \left({t\over 4C}+{C\over 6t}\right) {1\over a} -{t^2\over 2}\ln{a}
\eeq
\beq
F^{(1)} = -{1\over 24}\,\ln{\left(1-2 a\sqrt{1-a^2}\over a^2\right)}
\eeq
The radius of convergence of $F^{(g)}$ is $|T|< {1\over 6\sqrt{3}}$ for all $g$.

\subsubsection{Example square maps}

Consider the particular case $V(x)={t_4\over 4}x^4$, and write $T={t\,t_4\over C^2}$.
Define:
\beq
b=\sqrt{1-12 T}
\eeq

We find
\beq
\alpha=0 \virg \gamma^2 = {2 t\over C (1+b)}
\eeq
\beq
v_1={t\over \gamma} \virg v_2=0 \virg v_3 = -t_4 \gamma^3
\eeq

We find:
\beq
F^{(0)} = {t^2\over 2}\,\left( -{(1-b)^2\over 12 (1+b)^2} + {2(1-b)\over 3(1+b)} + \ln{\left({1+b\over 2}\right)}\right)
\eeq
\beq
F^{(1)} = -{1\over 12}\ln{\left(2b\over 1+b\right)}
\eeq
The radius of convergence of $F^{(g)}$ is $|T|< {1\over 12}$ for all $g$.

\subsection{2-Matrix model}

The 2-matrix model was introduced by Kazakov \cite{KazakovIsing}, as the Ising model on a  random map.
\beq
Z_{\rm formal} = {\int dM_1\, dM_2\, \ee{-{N\over t}\Tr(\VV_1(M_1)+\VV_2(M_2)-C_{12}\,M_1 M_2)}
\over \int dM_1\, dM_2\, \ee{-{N\over t}\Tr({C_{11}\over 2} M_1^2 + {C_{22}\over 2} M_2^2  - M_1 M_2)}}
\eeq
where $\VV_1(M_1)={C_{11}\over 2}M_1^2 -V_1(M_1)$ is a polynomial in $M_1$, such that $V_1(M_1)$ contains only terms with degree $\geq 3$, and
where $\VV_2(M_2)={C_{22}\over 2}M_2^2 -V_2(M_2)$ is a polynomial in $M_2$, such that $V_2(M_2)$ contains only terms with degree $\geq 3$, and we assume $C_{12}=1$:
\beq
C = \pmatrix{ C_{11} & -1 \cr -1 & C_{22}}
\eeq
Indeed, it generates surfaces made of polygons of two possible colors (call them polygons carrying a spin $+$ or $-$) glued by their sides (no crossing here).
The weight of each surface depends on the number of neighbouring polygons with same spin or different spin, which is precisely the Ising model.
If $C_{11}=C_{22}$ and $V_1=V_2$, this is an Ising model without magnetic field, otherwise it is an Ising model with magnetic field.

\medskip
Let us describe the solution.

Consider the following rational curve
\beq
\left\{
\begin{array}{l}
x(z)= \gamma z + \sum_{k=0}^{\deg V'_2} \alpha_k z^{-k} \cr
y(z)= \gamma z^{-1} + \sum_{k=0}^{\deg V'_1} \beta_k z^{k} \cr
\end{array}\right.
\eeq
where all coefficients $\gamma, \alpha_k, \beta_k$ are determined by:
\bea
y(z)-V'_1(x(z)) \mathop{\sim}_{z\to\infty} -{t\over \gamma z} + O(z^{-2}) \cr
x(z)-V'_2(y(z)) \mathop{\sim}_{z\to 0} -{tz\over \gamma} + O(z^{2}) \cr
\eea
The coefficients $\gamma, \alpha_k, \beta_k$ are algebraic functions of $t$, and we must choose the branch such that $\gamma\to 0$ at $t=0$.

The curve $(x(z),y(z))\, z\in\C$ is called the spectral curve. It is a genus zero algebraic curve.
There are $\deg V_2$  branch points solutions of $x'(z)=0$.

Notice that we have:
\beq
\Res_{z\to\infty} y dx  = t = - \Res_{z\to 0} y dx
\eeq

Then one has \cite{marcoF}:
\bea
F^{(0)}&=&{1\over 2}\Big( \Res_{z\to\infty} \VV_1(x) y dx + \Res_{z\to 0} (xy-\VV_2(y)) ydx \cr
&&  - t \Res_{z\to\infty} \VV_1(x) {dz\over z} - t \Res_{z\to 0} (xy-\VV_2(y)) {dz\over z}-{3\over 2}t^2 - t^2 \ln{\left(\gamma^2 \det C\over t\right)}
   \Big) \cr
\eea
and \cite{EKK}:
\beq
F^{(1)}= -{1\over 24}\ln{\left( {(\td{t}_{\deg(V_2)})^2\over t^2}\,\prod_{i=1}^{\deg V_2} \gamma y'(a_i)  \right)}
\eeq
where $a_i$ are the solutions of $x'(a_i)=0$, and $\td{t}_{\deg(V_2)}$ is the leading coefficient of $\VV_2$.

\medskip

The other $F^{(g)}$'s are found as follows \cite{CEO}:

Let $a_i$, $i=1,\dots,\deg V_2$ be the branch points, i.e. the solutions of $x'(a_i)=0$.
If $z$ is close to a branch-point $a_i$, we denote $\ovl{z}$ the other point such that
\beq
z\to a_i\virg
x(\ovl{z})=x(z) \quad {\rm and}\quad \ovl{z}\to a_i
\eeq
Notice that $\ovl{z}$ depends on the branch-point, i.e. $\ovl{z}$ is not globally defined.
We also define:
\beq
\Phi(z) = \int^z_{z_0} y dx
\eeq
$\Phi(z)$ depends on the base-point $z_0$ and on the path between $z$ and $z_0$, but the calculations below don't depend on that.

\smallskip

We define recursively:
\beq
W_{1}^{(0)}(p) =0
\eeq
\beq
W_{2}^{(0)}(p,q) ={1\over (p-q)^2}
\eeq
\bea
&& W_{k+1}^{(g)}(p,p_1,\dots,p_k) \cr
&=&  -{1\over 2}\,\sum_i \Res_{z\to a_i} {(z-\ovl{z})dz \over (p-z)(p-\ovl{z})(y(z)-y(\ovl{z}) x'(\ovl{z})}\,\Big( \cr
&& W_{k+2}^{(g-1)}(z,\ovl{z},p_1,\dots,p_k)
 + \sum_{h=0}^{g} \sum_{I\subset \{1,2,\dots,k\}} W_{1+|I|}^{(h)}(z,p_I) W_{1+k-|I|}^{(g-h)}(\ovl{z},p_{K/\{ I\} }) \Big) \cr
\eea
This system is a triangular system, and all $W_{k}^{(g)}$ are well defined in a finite number of steps $\leq {g(g+1)\over 2}+k$.

Then we have \cite{CEO}:
\beq
F^{(g)} =  {1\over 2-2g} \sum_i \Res_{z\to a_i} \Phi(z) W_{1}^{(g)}(z) \, dz \qquad , \,\, g>2
\eeq

\medskip
The 1-matrix case is a special case of this when the curve is hyperelliptical (in that case $\ovl{z}$ is globally defined), it corresponds to $\deg V_2=2$.

\subsection{Chain of matrices}

\beq
Z_{\rm formal} = {\int dM_1\,\dots  dM_p\, \ee{-{N\over t}\Tr(\sum_{i=1}^p \VV_i(M_i) - \sum_{i=1}^{p-1} M_i M_{i+1})}
\over \int dM_1\,\dots  dM_p\, \ee{-{N\over t}\Tr(\sum_{i=1}^p {C_{ii}\over 2} M_i^2 - \sum_{i=1}^{p-1} M_i M_{i+1})}}
\eeq
where $\VV_i(M_i)={C_{ii}\over 2}M_i^2 -V_i(M_i)$ is a polynomial in $M_i$, such that $V_i(M_i)$ contains only terms with degree $\geq 3$.
The matrix $C$ is:
\beq
C=\pmatrix{
C_{11} & -1 & & & & \cr
 -1 & C_{22} & -1 &&& \cr
& & &\ddots & & \cr
  & & & -1 & C_{pp} & -1 \cr
}
\eeq

Consider the following rational curve
\beq
x_i(z)= \sum_{k=-s_i}^{r_i} \alpha_{i,k} z^k \qquad \qquad \forall i=0,\dots , p+1
\eeq
where all coefficients $\gamma, \alpha_k, \beta_k$ are determined by:
\bea
&& x_{i+1}+x_{i-1} = \VV'(x_i)  \qquad \qquad \forall i=1,\dots , p \cr
&& x_0(z) \mathop{\sim}_{z\to\infty} {t\over \gamma z} + O(z^{-2}) \cr
&& x_{p+1}(z) \mathop{\sim}_{z\to 0} {tz\over \gamma} + O(z^{2}) \cr
\eea
The coefficients $\gamma, \alpha_{i,k}$ are algebraic functions of $t$, and we must choose the branch such that $\gamma\to 0$ at $t=0$.

The curve $(x_1(z),x_2(z))\, z\in\C$ is called the spectral curve. It is a genus zero algebraic curve.

Notice that we have $\forall i=1,\dots, p-1$:
\beq
\Res_{z\to\infty} x_{i+1} dx_i  = t = - \Res_{z\to 0} x_{i+1} dx_i
\eeq

Then one has \cite{eynmultimat}:
\bea
F^{(0)}&=&{1\over 2}\Big( \sum_{i=1}^p \Res_{z\to\infty} (\VV_i(x_i) -{1\over 2}x_i \VV'_i(x_i)) x_{i+1} dx_i \cr
&&  - t \sum_{i=1}^p \Res_{z\to\infty} (\VV_i(x_i) -{1\over 2}x_i \VV'_i(x_i)) {dz\over z}  -  t^2 \ln{\left({\gamma^2\det C\over t}\right)}
  \Big) \cr
\eea
$F^{(1)}$ and the other $F^{(g)}$'s have never been computed, but it is strongly believed that they should be given by the same formula as in the 2-matrix case.

\subsection{Closed Chain of matrices}

\beq
Z_{\rm formal} = {\int dM_1\,\dots  dM_p\, \ee{-{N\over t}\Tr(\sum_{i=1}^p \VV_i(M_i) - \sum_{i=1}^{p-1} M_i M_{i+1} - M_p M_1)}
\over \int dM_1\,\dots  dM_p\, \ee{-{N\over t}\Tr(\sum_{i=1}^p {C_{ii}\over 2} M_i^2 - \sum_{i=1}^{p-1} M_i M_{i+1} - M_p M_1)}}
\eeq
It is the case where the matrix $C$ of quadratic interactions has the form:
\beq
C=\pmatrix{
C_{11} & -1 & & & & -1 \cr
 -1 & C_{22} & -1 &&& \cr
& & &\ddots & & \cr
-1  & & & -1 & C_{pp} & -1 \cr
}
\eeq
This model is yet unsolved, apart from very few cases ($p=2$, $p=3$  (Potts Model), $p=4$ with cubic potentials), and there are also results in the large $p$ limit.
This model is still a challenge.

\subsection{O(n) Model}

\beq
Z_{\rm formal} = {\int dM\,dM_1\,\dots  dM_n\, \ee{-{N\over t}\Tr( {C_M\over 2}M^2 + {C\over 2} \sum_{i=1}^n M_i^2 \, - V(M)  - \sum_{i=1}^{n} M M_i^2)}
\over \int dM\,dM_1\,\dots  dM_n\, \ee{-{N\over t}\Tr( {C_M\over 2}M^2 + {C\over 2} \sum_{i=1}^n M_i^2)}}
\eeq
where $V$ contains at least cubic terms.

We write:
\beq
{\cal V}(x) = -V(-(x+{C\over 2})) + {C_M\over 2}(x+{C\over 2})^2
\eeq

\smallskip

This model can easily be analytically continued for non integer $n$'s. Indeed, combinatoricaly, it is the generating function of a loop gas living on the maps.
$n$ is  the ``fugacity'' of loops, i.e. the $n$ dependence of the weight of each configuration is $n^{\hbox{\# loops}}$, and the $C$ dependence is $C^{-\hbox{length of  loops}}$.
One often writes:
\beq
n= 2\cos{(\nu\pi)}
\eeq
The $O(n)$ model was introduced by I. Kostov in \cite{KostovOn} then in \cite{KostovOnADE, KostovDuplantier}, and it
 plays a very important role in many areas of physics, and lately, it has started to play a special role in string theory, as an effective model for the check of ADS-CFT correspondence.

\medskip

The leading order 1-cut solution of this model is well known \cite{EOnZinn, EKOn1, EKOn2}.

For any two distinct complex numbers $a$ and $b$, define:
\beq
 m=1-{a^2\over b^2} \virg \tau = {iK'(m)\over K(m)}
\eeq
where $K(m)=K'(1-m)$ are the complete elliptical integrals (\cite{eliptical}).

We also consider the following elliptical function (defined on the torus $(1,\tau)$):
\beq
x(u)= i b\,\,{\cn(2K(m)u,m)\over \sn(2K(m)u,m)}
\eeq
where $\sn$ and $\cn$ are the elliptical sine and cosine functions.

Then we define the following function on the universal covering (i.e. on the complex plane):
\beq
G_\nu(u) = H(\ee{i\nu\pi/2}\,\, {\theta_1(u+{\nu\over 2})\over \theta_1(u)} + \ee{-i\nu\pi/2}\,\, {\theta_1(u-{\nu\over 2})\over \theta_1(u)})
\eeq
where $H$ is a normalization constant such that:
\beq
\mathop{{\rm lim}}_{u\to 0} G_\nu(u)/x(u) =  1
\eeq

It satisfies:
\beq
G_\nu(u+\tau)+G_\nu(u-\tau)-nG_\nu(u)=0 \virg G(u+1)=G(u) \virg G(u)=G(-\tau-u)
\eeq
We have:
\beq
G_\nu(u)^2 + G_\nu(-u)^2 - n G_\nu(u)G_\nu(-u) = (2+n) (x^2(u)-e_\nu^2)
\virg
e_\nu = x({\nu\over 2})
\eeq
\beq
G_{1-\nu}(u)^2 + G_{1-\nu}(-u)^2 + n G_{1-\nu}(u)G_{1-\nu}(-u) = (2-n) (x^2(u)-e_{1-\nu}^2)
\eeq
\bea
&& G_{1-\nu}(u)G_\nu(u) - G_{1-\nu}(-u)G_{\nu}(-u) + {n\over 2} G_{1-\nu}(-u)G_{\nu}(u) -{n\over 2} G_{1-\nu}(u)G_{\nu}(-u) \cr
&=& x(u)\, b {2\sin{(\nu\pi)}\over \sn{(\nu K)}\,\sn{((1-\nu)K)}}
\eea

Then we define:
\beq
A(x^2)={\rm Pol}\Big( { (2+n)(x^2-e_\nu^2) ({\cal V}'.G_{1-\nu})_+ - x b c ({\cal V}'.G_\nu)_-  \over (x^2-b^2)^2}  \Big)
\eeq
\beq
B(x^2)={\rm Pol}\Big( { (2-n)(x^2-e_{1-\nu}^2) {1\over x}\,({\cal V}'.G_{\nu})_-  -  b c ({\cal V}'.G_{1-\nu})_+  \over (x^2-b^2)^2}  \Big)
\eeq
where ${\rm Pol}$ means the polynomial part in $x$ at large $x$ (i.e. at $u\to 0$), and the subscript $+$ and $-$ mean the even and odd part, and
where
\beq
c={2\sin{(\nu\pi)}\over \sn{(\nu K)}\,\sn{((1-\nu)K)}}
\eeq
$A$ and $B$ are polynomials of $x^2$.

Then, $a$ and $b$ are completely determined by the 2 conditions:
\beq
\Res_{u\to 0} \left({A(x^2(u)) \, G_{1-\nu}(u)_+\over x(u)} + B(x^2(u)) \, G_{\nu}(u)_- \right) \, dx(u) = 0
\eeq
\beq
\Res_{u\to 0} \left(A(x^2(u)) \, G_{1-\nu}(u)_- + x(u) B(x^2(u)) \, G_{\nu}(u)_+\right)\,dx(u) = t
\eeq

Once we have determined $a$, $b$, $m$, $A$, $B$, we define the resolvent:
\beq
\om(u) = A(x(u)^2) G_{1-\nu}(u) + x(u) B(x(u)^2) G_{\nu}(u)
\eeq
which is the first term of the formal large $N$ expansion of:
\beq
{1\over N}\left<\Tr{1\over x(u)-M}\right> = \om(u) + O(1/N^2)
\eeq
and which satisfies:
\beq
\om(u+\tau)+\om(u-\tau)+n\om(u) = {\cal V}'(x(u))
\virg \om(u+1)=\om(u) \virg \om(u)=\om(-\tau-u)
\eeq

The free energy is then found by:
\beq
{\partial F^{(0)}\over \partial t^2} = (1-{n\over 2}) \ln{(a^2\,g(m))}
\eeq
where
\beq
{g'\over g} = {1\over m(1-m)\sn^2(\nu  K(m),m)}
\eeq

All this is  described in \cite{eynardthese}.
The special cases of $n$ integer, and in general when $\nu$ is rational, can be computed with more details.

\subsection{Potts model}

\beq
Z_{\rm formal} = {\int dM_1\,\dots  dM_Q\, \ee{-{N\over t}\Tr(  \sum_{i=1}^{Q} {1\over 2} M_i^2 + V(M_i) + {C\over 2} \sum_{i,j}^Q M_i M_j)}
\over \int dM\,dM_1\,\dots  dM_n\, \ee{-{N\over t}\Tr( \sum_{i=1}^{Q} {1\over 2}M_i^2 + {C\over 2} \sum_{i,j}^Q M_i M_j)}}
\eeq
or equivalently:
\beq
Z_{\rm formal} = {\int dM\, dM_1\,\dots  dM_Q\, \ee{-{N\over t}\Tr(  \sum_{i=1}^{Q} {1\over 2} M_i^2 + V(M_i) + {C\over 2} M^2 -  C \sum_{i=1}^Q M M_i )}
\over \int dM\, dM_1\,\dots  dM_Q\, \ee{-{N\over t}\Tr(  \sum_{i=1}^{Q} {1\over 2} M_i^2  + {C\over 2} M^2 -  C \sum_{i=1}^Q M M_i )}}
\eeq

The Random lattice Potts model was first introduced by V. Kazakov in \cite{KazakovPotts, KazakovPottsbis}.
The loop equations have been written and solved to leading order, in particular in \cite{Daul, PZinnPotts, BEPotts}.

\subsection{3-color model}

\beq
Z_{\rm formal} = {\int dM_1\,dM_2\,dM_3\, \ee{-{N\over t}\Tr( {1\over 2}(M_1^2+M_2^2+M_3^2) - g (M_1 M_2 M_3+M_1 M_3 M_2))}
\over \int dM_1\,dM_2\,dM_3\, \ee{-{N\over t}\Tr( {1\over 2}(M_1^2+M_2^2+M_3^2))}}
\eeq
The loop equations have been written and solved to leading order, in particular in \cite{KE3coul, Kostov3coul}.

\subsection{6-vertex model}

\beq
Z_{\rm formal} = {\int dM\,dM^\dagger\, \ee{-{N\over t}\Tr( M M^\dagger - M^2 M^{2\dagger} + \cos\beta (M M^\dagger)^2)}
\over \int dM\,dM^\dagger\, \ee{-{N\over t}\Tr( M M^\dagger)}}
\eeq
where $M$ is a complex matrix.
The loop equations have been written and solved to leading order, in particular in \cite{PZinn6V, Kostov6V}.

\subsection{ADE models}

Given a Dynkin diagram of A, D or E Lie algebra, and let $A$ be its adjacency matrix ($A_{ij}=A_{ji}=$\# links between $i$ and $j$).
We define:
\beq
Z_{\rm formal} = \int \prod_i dM_i \prod_{i<j} dB_{ij}\,\,\ee{-{N\over T}\Tr(\sum_i {1\over 2}M_i^2 -{g\over 3} M_i^3 + {1\over 2}\sum_{i,j} B_{ij}B_{ij}^t+ {K\over 2}\sum_{i,j} A_{ij} B_{ij}B_{ij}^t M_i )}
\eeq
where $B_{ji}=B_{ij}^t$ are complex matrices, and $M_i$ are hermitian matrices.

The loop equations have been written and solved to leading order, in particular in \cite{KostovOnADE, KostovADE}.

\subsection{ABAB models}

\beq
Z_{\rm formal} =
\int \prod_{i=1}^{n_1} dA_i \prod_{j=1}^{n_2} dB_{j}\,\,\ee{-{N\over T}\Tr(\sum_i {A_i^2\over 2}+\sum_i {B_i^2\over 2} + g  \sum_{i,j} A_i B_j A_i B_j}
\eeq
This model is yet unsolved, apart from few very special cases \cite{KPzinnABAB}.
However, its solution would be of great interest in the understanding of Temperley Lieb algebra.

\section{Discussion}

\subsection{Summary of some known results}

We list here some properties which are rather well understood.

\begin{itemize}

\item The fact that formal matrix integrals are generating functions for counting discrete surfaces (also called maps) is well understood, as was explained in this review.

\item The fact that formal matrix integrals i.e. generating functions of maps satisfy Schwinger--Dyson equations is well understood.

\item The fact that formal matrix integrals and convergent integrals don't coincide in general is well understood.
In the examples of the 1-matrix model, 2-matrix Ising model or chain of matrices, it is understood that they may coincide only in the ``1-cut case'', i.e. if the classical spectral curve has genus zero.

\item The fact that the formal 1-matrix, 2-matrix or chain of matrices integrals are $\tau$ functions of some integrable hierarchies is well understood too.

\item For the formal 1-matrix model and 2-matrix Ising model, all $F^{(g)}$'s have been computed explicitly. The result is written in terms of residues of rational functions.

\item For the chain of matrices, $F^{(0)}$ is  known explicitly \cite{eynmultimat}, and it is strongly believed that all $F^{(g)}$'s are given by the same expression as for the 2-matrix model.

\item Multi-cut formal matrix models are well studied too, and they can be rewritten in terms of multi-matrix models. For the 1 and 2 matrix models, the expressions of the $F^{(g)}$'s are known explicitly (in terms of residues of meromorphic forms on higher genus spectral curves).

\end{itemize}

\subsection{Some open problems}

\begin{itemize}

\item the large $N$ asymptotics of convergent matrix integrals, are not directly relevant for combinatorics, but are a very important question for other applications, in particular bi-orthogonal polynomials,
universal statistical properties of eigenvalues of large random matrices, and many other applications.
This question is highly non-trivial, and so far, it has been solved only for the 1-matrix model \cite{BlIt,dkmvz,EML}, and a  few other cases (e.g. the 6-vertex model \cite{Bleher6V}).
The method mostly used to prove the asymptotic formulae is the Riemann-Hilbert method \cite{BlIt, dkmvz}, which consists in finding a Riemann-Hilbert problem for the actual matrix integral and for its conjectured asymptotics, and compare both.
There is a hope that probabilists' methods like large deviations  could be at leats as efficient.

\item For the 2-matrix model and chain of matrices, the topological expansion of formal ``mixed'' expectation values (e.v. of invariant monomials whose words contain at least 2 different letters) has not yet been computed.
This problem is a challenge in itself, and has applications to the understanding of ``boundary conformal field theory''. In terms of combinatorics, it corresponds to find generating functions for open surfaces with boundaries of prescribed colors.

\item Many matrix models have not yet been solved, even to planar order. For instance the closed chain of matrices where $C_{ij} = \delta_{i,j+1}+\delta_{i,j-1}+C_i \delta_{ii}$ and $C_{p1}=C_{1p}=1$.
For the Potts model, 6-vertex model, 3-color model, O(n) model, ADE models, only the planar resolvent is known.
For the $A_nB_mA_nB_m$, almost nothing is known, although this model describes the combinatorics of Temperly Lieb algebra.

\item Limits of critical points are still to be understood and classified. Only a small number of critical points have been studied so far. They have been related to KdV or KP hierarchies.
Critical points, i.e. radius of convergence of the $t$-formal power series, are in relationship with asymptotics numbers of large maps (through Borel's transform), and thus critical points allow to count maps made of a very large number of polygons, i.e. they can be interpreted as counting continuous surfaces (this was the idea of 2D quantum gravity in the 80's and early 90's).
This is yet to be better understood \cite{Cicuta, BlEycrit}.

\item extensions to other types of matrices (non-hermitian) have been very little studied compared  to the hermitian case, and much is still to be understood.
For instance real symmetric matrices or quaternion-self-dual matrices have been studied from the begining \cite{Mehta}, and they count non-orientable maps.
Complex matrices, and normal complex matrices have played an increasing role in physics, because of their relationship with Laplacian growth problems \cite{WiegZab}, or some limits of string theory \cite{bmn}.
Complex matrices count maps with arrows on the edges (see the 6-vertex model \cite{PZinn6V}).
Other types of matrix ensembles have been introduced in relationship with symmetric spaces \cite{zirnbauer}, and it is not clear what they count.

\item And there are so many applications of random matrices to physics, mathematics, biology, economics, to be investigated...

\end{itemize}

\subsection*{Acknowledgements}

I would like to thank Philippe Di Francesco for his critical reading of the manuscript. This work is partly supported by the Enigma European network MRT-CT-2004-5652, by the ANR project G\'eom\'etrie et int\'egrabilit\'e en physique math\'ematique ANR-05-BLAN-0029-01,
and by the Enrage European network MRTN-CT-2004-005616.


\begin{thebibliography}{99}

\bibitem{eliptical} M. Abramowitz, I. A. Stegun,`` Handbook of Mathematical Functions'', (1964) Dover Publications, New York. ISBN 0-486-61272-4.

\bibitem{marcoF} M. Bertola, ''Second and Third Order Observables of the Two-Matrix Model'',
JHEP 0311 (2003) 062, hep-th/0309192.

\bibitem{marcomo} M. Bertola, Mo Man Yue,

\bibitem{bmn} D. Berenstein, J.~M. Maldacena, H. Nastase, JHEP {\bf 04},  013  (2002).

\bibitem{BI} P.M. Bleher and A.R. Its, eds.,  ``Random Matrix Models and
Their Applications'', {\em MSRI Research Publications} {\bf 40}, Cambridge Univ.
Press, (Cambridge, 2001).

\bibitem{BlIt} P. Bleher, A. Its, ``Semiclassical asymptotics of
orthogonal polynomials, Riemann-Hilbert problem,
and universality in the matrix model'' {\em Ann. of Math.}
 (2) {\bf 150}, no. 1, 185--266 (1999).

\bibitem{BlEycrit} P. Bleher, B. Eynard,
''Double scaling limit in random matrix models and a non-linear hierarchy of differential equations'',
 J. Phys. A36 (2003) 3085-3106, xxx, hep-th/0209087.

\bibitem{Bleher6V} P. Bleher, V. Fokin, ``Exact Solution of the Six-Vertex Model with Domain Wall Boundary Conditions. Disordered Phase'', math-ph/0510033, to appear in CMP.

\bibitem{BDE}
G. Bonnet, F. David, B. Eynard,
Breakdown of universality in multi-cut matrix models,
J.Phys. A: Math. Gen. {\bf 33} (2000) 6739-6768.

\bibitem{BEPotts} G. Bonnet, B. Eynard, ''The Potts-q random matrix model : loop equations, critical exponents, and rational case'', Phys.Lett. B463 (1999) 273-279, hep-th/9906130.

\bibitem{BrezinDeo}
E. Br\'ezin, N. Deo, Phys.  Rev. E 59 (1999) 3901-3910,
cond-mat/98050096.

\bibitem{BIPZ} E. Brezin, C. Itzykson, G. Parisi, and J. Zuber, Comm. Math. Phys. {\bf 59},   35  (1978).

\bibitem{chekhovF1} L.Chekhov, "Genus one corrections to multi-cut matrix model solutions", Theor. Math. Phys. 141 (2004) 1640-1653, hep-th/0401089.

\bibitem{eynchekhov} L. Chekhov, B. Eynard, ''Hermitean matrix model free energy: Feynman graph technique for all genera'',
JHEP009P0206/5, hep-th/0504116.

\bibitem{CEO} L. Chekhov, B. Eynard, N. Orantin, ''Free energy topological expansion for the 2-matrix model'',
math-ph/0603003.

\bibitem{Cicuta} G.M.Cicuta ``Phase transitions and random matrices'', cond-mat/0012078.

\bibitem{Daul}  J.M. Daul hep-th/9502014.

\bibitem{Matrixsurf} F. David, ``Planar diagrams, two-dimensional lattice gravity and surface models'', {\em Nucl. Phys.} {\bf B 257 [FS14]} 45 (1985).

\bibitem{dkmvz} P. Deift, T. Kriecherbauer, K. T. R. McLaughlin,
S. Venakides, Z. Zhou, ``Uniform asymptotics for polynomials
orthogonal with respect to varying exponential weights and
applications to universality questions in random matrix theory'', {\it
Commun. Pure Appl. Math.} {\bf 52}, 1335--1425 (1999).

\bibitem{DGZ} P. Di Francesco, P. Ginsparg, J. Zinn-Justin, ``2D Gravity and
Random Matrices'', {\em Phys. Rep.} {\bf 254}, 1 (1995).

%\bibitem{Dijgrafvafa} R. Dijkgraaf, C. Vafa, ''A Perturbative Window into Non-Perturbative Physics'', hep-th/0208048,

\bibitem{KostovDuplantier} B. Duplantier and I. Kostov, Phys. Rev. Lett. 61 (1988) 1433.

\bibitem{EML} N. Ercolani K. Mac Laughlin, ''Asymptotics of the partition function for random matrices via Riemann--Hilber technics and applications to graphical enumeration'',
IMRN 22003, no 14, 755-820.

\bibitem{eynloop1mat} B. Eynard, ''Topological expansion for the 1-hermitian matrix model correlation functions''
JHEP: JHEP/024A/0904, hep-th/0407261.

\bibitem{EKK} B. Eynard, D. Korotkin, A. Kokotov, ''Genus one contribution to free energy in hermitian two-matrix model'',
Nucl.Phys. B694 (2004) 443-472, hep-th/0403072.

\bibitem{eynmultimat} B. Eynard, ''Master loop equations, free energy and correlations for the chain of matrices'', JHEP11(2003)018, hep-th/0309036.

\bibitem{EOnZinn} B. Eynard, J. Zinn-Justin, ''The O(n) model on a random surface: critical points and large order behaviour'', Nuc. Phys. B386 558-591, (1992),
hep-th/9204082.

\bibitem{EKOn1} B. Eynard, C. Kristjansen, ''Exact solution of the O(n) model on a random lattice'', Nuc. Phys. B455 577-618, (1995),
hep-th/9506193.

\bibitem{EKOn2} B. Eynard, C. Kristjansen, ''More on the exact solution of the O(n) model on a random lattice and an investigation of the case |n|>2'', Nuc. Phys. B466 463-487, (1996),
hep-th/9512052.

\bibitem{KE3coul} B. Eynard, C. Kristjansen, '' An iterative solution of the three colour Problem on a random lattice'', Nuclear Physics B516 [FS] (1998) 529,
cond-mat/9710199.

\bibitem{eynardthese} B. Eynard, ``Gravitation quantique bidimensionnelle et matrices aléatoires'', SPHT 95/033, phd thesis, Université Paris 6 (1995), http://www-spht.cea.fr/articles/t95/033/.

\bibitem{eynhabilit} B. Eynard, "The 2-matrix model, biorthogonal polynomials, Riemann-Hilbert problem, and algebraic geometry", Habilitation thesis, Paris VII university, in french, math-ph/0504034.

\bibitem{gross:1991}
{\em Two dimensional quantum gravity and random surfaces}, edited by D. Gross,
  T. Piran, and S. Weinberg (Jerusalem winter school, World Scientific, .,
  1991).

\bibitem{Guhr} T. Guhr, A. Mueller-Groeling, H.A. Weidenmuller, ``Random
matrix theories in quantum physics: Common concepts'', {\em Phys. Rep.}
{\bf 299}, 189 (1998).

\bibitem{guionet} A. Guionnet, O. Zeitouni, ``Large deviations asymptotics for spherical integrals'',
{\em J. F. A.} {\bf 188}, 461--515 (2002).

\bibitem{guionnet2} A. Guionnet, E. Segala Maurel, ``Combinatorial aspects of matrix models''.

\bibitem{KazakovIsing} V.A. Kazakov, ``Ising model on a dynamical planar random
lattice: exact solution'', {\em Phys Lett.} {\bf A119}, 140-144 (1986).

\bibitem{KazakovGQ} V.A. Kazakov , "Bilocal regularization of models of random surfaces", Phys.Lett.B150:282-284,1985.

\bibitem{KazakovGQbis} V.A. Kazakov, A. A. Migdal, I.K. Kostov, "Critical Properties Of Randomly Triangulated Planar Random Surfaces", Phys.Lett.B157:295-300,1985.

\bibitem{KazakovPotts} V.A. Kazakov, "Exactly solvable Potts models, bond- and tree-like percolation on dynamical (random) planar lattice", Nuclear Physics B Proceedings Supplements, Volume 4, p. 93-97.

\bibitem{KazakovPottsbis} V.A. Kazakov, "Percolation On A Fractal With The Statistics Of Planar Feynman Graphs: Exact Solution", Mod.Phys.Lett.A4:1691,1989.

\bibitem{KPzinnABAB} V. Kazakov, P. Zinn-Justin, ''Two-Matrix model with ABAB interaction'', Nuclear Physics B 546 647 (1999).

\bibitem{Kanz} E. Kanzieper and V. Freilikher, Phys. Rev. E 57, 6604 (1998), cond-mat/9709309.

\bibitem{KostovOn} I.K. Kostov, "O(n) vector model on a planar random lattice: spectrum of anomalous dimensions", Mod.Phys.Lett.A4:217,1989.

\bibitem{KostovOnADE} I. Kostov, "Strings with Discrete Target Space", Nucl.Phys. B376 (1992) 539-598, hep-th/9112059.

\bibitem{Kostov3coul} I.  Kostov, ''Exact Solution of the Three-color Problem on a Random Lattice'', Physics Letters B 549 245 (2002).

\bibitem{Kostov6V} I. Kostov, ''Exact Solution of the Six-Vertex Model on a Random Lattice'', Nuclear Physics B 575 513 (2000).

\bibitem{KostovADE} I. Kostov, ''Gauge Invariant Matrix Model for the ADE Closed Strings'', Physics Letters B 297 74 (1992).

\bibitem{Mehta} M.L. Mehta, {\em Random Matrices}, 2nd edition, (Academic Press, New York, 1991).

\bibitem{thooft:1974} G. 't~Hooft, Nuc.Phys. {\bf B72},  461  (1974).

\bibitem{Tutte1} W.T. Tutte, ''A census of planar triangulations'', {\em Can. J. Math.} {\bf 14} (1962) 21-38.

\bibitem{Tutte2} W.T. Tutte, ''A census of planar maps'', {\em Can. J. Math.} {\bf 15} (1963) 249-271.

\bibitem{Moerbeke:2000} P. Van Moerbeke, Random Matrices and their applications, MSRI-publications {\bf 40},  4986  (2000).

\bibitem{Wadia} Spenta R. Wadia, "On The Dyson-Schwinger Equations Approach To The Large N Limit: Model Systems And String Representation Of Yang-Mills Theory", Phys.Rev.D24:970,1981.

\bibitem{WiegZab} P. Wiegmann, A. Zabrodin, ``Large N expansion for normal and complex matrix ensembles'', hep-th/0309253.

\bibitem{Wigner} E.P. Wigner, {\em Proc. Cambridge Philos. Soc.} {\bf 47}, 790 (1951).

\bibitem{PZinnPotts} P. Zinn-Justin, ''The dilute Potts model on random surfaces'', Journal of Statistical Physics 98 245 (2001).

\bibitem{PZinn6V} P. Zinn-Justin, ''The six-vertex model on random lattices'', Europhysics Letters 50 15 (2000).

\bibitem{zirnbauer} M. R. Zirnbauer, ``Riemannian symmetric superspaces and their origin in random matrix theory'', J. Math. Phys. 37 (1996) 4986, math-ph/9808012.

\end{thebibliography}
\end{document}